\documentclass[draft]{agujournal2019}
\usepackage{url} 
\usepackage{lineno}
\usepackage[inline]{trackchanges} 
\usepackage{soul}

\usepackage{float}
\usepackage{xcolor}
\usepackage{graphicx}
\usepackage{caption}
\usepackage{dcolumn}
\usepackage{bm}
\usepackage{amsmath}
\usepackage{amssymb}
\usepackage{cleveref}
\usepackage{amsmath}
\usepackage{booktabs}
\usepackage{multirow}

\draftfalse

\journalname{Applied Computing \& Geosciences}

\begin{document}



%
%







\correspondingauthor{J. G.-S.}{joaquin.garciasuare@epfl.ch}

\begin{center}
    \large \textbf{Data-driven Dynamic Friction Models\\ based on Recurrent Neural Networks} \\
    \vspace{0.5cm}
    \normalsize 
    \textbf{\color{black} Gaëtan Cortes}, 
    \textbf{Joaquin Garcia-Suarez} \\
    \textit{ 
Institute of Civil Engineering,
            \'{E}cole Polytechnique F\'{e}d\'{e}rale de Lausanne (EPFL),\\ CH 1015 Lausanne, 
            Switzerland} \\
\texttt{\color{black} gaetan.cortes@epfl.ch}, 
    \texttt{joaquin.garciasuarez@epfl.ch} \\
\end{center}

\section*{Abstract}
In this concise contribution, it is demonstrated that Recurrent Neural Networks (RNNs) based on Gated Recurrent Unit (GRU) architecture, possess the capability to learn the complex dynamics of rate-and-state friction (RSF) laws from synthetic data.
The data employed for training the network is generated through the application of traditional RSF equations coupled with either the aging law or the slip law for state evolution. 
A novel aspect of this approach is the formulation of a loss function that explicitly accounts for the direct effect by means of automatic differentiation. 
It is found that the GRU-based RNNs effectively learns to predict changes in the friction coefficient resulting from velocity jumps (with and without noise in the target data), 
thereby showcasing the potential of machine learning models in capturing and simulating the physics of frictional processes. Current limitations and challenges are discussed.

\noindent \textbf{Keywords}: recurrent neural networks, geophysics, friction, constitutive models, data 

\noindent \textbf{Code/Data Link}: \texttt{zenodo.org/records/13341474}

\section{Introduction}
Understanding the dynamics of friction is paramount across several disciplines, including mechanical engineering, civil engineering, and geophysics \cite{Rabinowicz:1951,Scholz:book}. 
%
%
Friction dynamics play a critical role in earthquake mechanics, influencing the initiation, propagation, and arrest of seismic slips along faults \cite{Marone:1998}. 
Numerous models have been proposed to describe frictional behavior, from classical laws like Amontons-Coulomb friction \cite{Scholz:book}, which simplifies friction as a constant proportion of the normal load, to much more complex ones. However, these models often fall short in capturing the full range of observed behaviors \cite{Marone:1998,Woodhouse:2015}. 
Rate-and-state friction (RSF) laws stand out by explicitly addressing the velocity and time-dependent effects, thereby providing a more comprehensive framework for understanding the seismic cycle \cite{Rice:1996}. A customary expression of this law is \cite{Marone:1998}:

\begin{align}
\label{eq:rsf_laws}
    \mu 
    = 
    \mu_0 
    + 
    a \ln\left(\frac{V}{V_{ref}}\right) 
    + 
    b \ln\left(\frac{V_{ref} \theta}{D_c}\right) \, ,
\end{align}

where $\mu$ is the friction coefficient, $\mu_0$ is the friction coefficient at a reference velocity $V_{ref}$, $V$ is the sliding velocity, and $a$ is a positive constant representing the direct effect's sensitivity to velocity changes \cite{Dieterich:1979}, $b$ is a constant that, along with $a$, determines the frictional response to changes in sliding velocity and state evolution \cite{Ruina:1983}. 

Friction modeling \cite{Johnson:1985,Bizzarri:2010} aims to capture essential features observed in experimental settings \cite{Beeler:1996}, notably the direct effect and healing \cite{Scholz:book}. 
Healing, also referred to as state evolution effect, reflects the time-dependent increase in frictional resistance during stationary contact, attributed to the contact area's growth at asperity contacts or chemical bonding at the interface \cite{Scholz:book}. 
This process is often modeled by the evolution of an internal variable $\theta$ (termed ``state''), with its simplest evolution equations being the ``aging law'' and the ``slip law'':

\begin{subequations}
\begin{align}
    \label{eq:state_evolution_aging}
    \dot{\theta} = {d \theta \over d t} &= 1 - \frac{\theta V}{D_c} \, , \\
    \label{eq:state_evolution_slip}
    \dot{\theta} = {d \theta \over d t} &= - \frac{\theta V}{D_c} \log \left(\frac{\theta V}{D_c}\right) \, ,
\end{align}
\label{eq:state_equations}
\end{subequations}

where $t$ is time and $D_c$ is a critical slip distance over which the state variable evolves significantly \cite{Marone:1993}. 
In the aging law, the state is interpreted as a measure of effective age of contacts between the asperities of the opposed surfaces \cite{Dieterich:1979,Dieterich:1994b}, a logical position given that contact area between two rough surfaces is also to grow logarithmically over time during periods of rest \cite{Berthoud:1999,Baumberger:2006}. 
Notwithstanding this intuitive inference, the slip law offers a picture that may be better backed by experiments \cite{Bhattacharya:2017,Bhattacharya:2022}, in which the sensitivity to sliding does not ever disappear, not even at tiny sliding rates \cite{Bhattacharya:2022}. 
%
%

Despite their advantages, RSF laws are not without limitations, e.g., 
identifying the parameters $a$, $b$, and $D_c$ from experimental data poses significant challenges due to the complex interplay between different physical processes at the frictional interface \cite{Marone:1998}. 
\textcolor{black}{Criticism has been levied against the use of the ``state" variable in RSF laws for its lack of a clear physical basis, leading to calls for models that include more explicit physical mechanisms \cite{Rice:1983, Lapusta:2000}. 
For example, the parameters that appear in \cref{eq:rsf_laws} are assumed to be constants, and they may not \cite{Noda:2013,Veedu:2016,Xu:2018,Urata:2018,Im:2020}. 
Furthermore, the need to regularize these laws to address numerical instabilities and ensure the physical consistency of simulations has been highlighted \cite{Rubin:2005, BarSinai:2012}. 
Recently, laboratory \cite{Bizzarri:2024} and virtual experiments \cite{Ferdowsi:2020,Ferdowsi:2021} have shown the limitations of RSF laws even in highly-controlled interface conditions. 
}
Given these challenges and the limitations of existing models, there is a clear need to explore new methodologies for understanding friction dynamics. 
This necessity is particularly acute for modeling phenomena that span multiple scales of time and space \cite{BenZion:2001}. 
Herein lies the potential of novel data-driven approaches, such as Recurrent Neural Networks (RNNs) \cite{Goodfellow:2016}, when it comes to overcome the limitations of traditional friction laws.

Neural networks have revolutionized the field of machine learning, offering powerful tools for modeling complex, non-linear relationships in data across a myriad of disciplines \cite{LeCun:2015}.
At their core, neural networks are composed of layers of interconnected nodes or``neurons", each capable of performing simple computations. 
Through the process of training, these networks adjust their internal parameters to minimize the difference between their output and the desired outcome, effectively learning to map inputs to outputs \cite{Hochreiter:1997}.
Among the various architectures of neural networks, Recurrent Neural Networks (RNNs) stand out for their ability to process sequential data. 
Unlike feedforward neural networks, RNNs possess a form of memory that allows them to incorporate information from previous inputs into the current processing step \cite{Cho:2014}. 
This characteristic makes RNNs particularly suited for tasks involving sequential data.
Two notable advancements in RNN architecture are the Gated Recurrent Unit (GRU) and Long Short-Term Memory (LSTM) networks. 
Both GRU and LSTM address a fundamental limitation of basic RNNs known as the vanishing gradient problem associated to recurrent connections \cite{Goodfellow:2016}, which makes it difficult for RNNs to learn dependencies between events that occur at long intervals in the input data \cite{Bengio:1994,Hochreiter:1997}. 
GRUs simplify the LSTM architecture while retaining its ability to capture long-term dependencies, making it both efficient and powerful for a wide range of tasks \cite{Cho:2014}.
The application of RNNs, GRUs, and LSTMs has been widespread and transformative across many fields, including physics \cite{PINNs}. 
Particularly, their success in modeling path-dependent plasticity of materials highlights their potential \cite{Bessa:2019,Frankel:2019,Sun:2020,Mohr:2020,Masi:2021,Rimoli:2021,Darve:2021,Chen:2022}. Path-dependent plasticity is a critical aspect in the study of materials science, where the history of material deformation affects its current and future state. 
RNNs have been shown to effectively model these complex relationships, capturing the history-dependent nature of material behavior \cite{Mohr:2021,Dornheim:2023}. 
This success in modeling path-dependent phenomena provides a strong foundation for the belief that similar methodologies can be applied to the study of friction. 
Friction, much like material plasticity, is inherently dependent on history, in its case of contact and slip between surfaces. 
%
%

The goal of this short communication is to show that RNNs, based on GRUs, can capture and predict evolution of interfaces \textit{unbeknownstly} governed by RSF laws based on either the aging law, \cref{eq:state_evolution_aging}, or the slip law \cref{eq:state_evolution_slip}. 
This approach is essentially different from the ones used recently both by Ishiyama and collaborators \cite{Ishiyama:2023}, who used machine learning -- random forest -- \cite{Breiman:2001} to improve the estimation of RSF parameters \cref{eq:rsf_laws}, 
and 
\textcolor{black}{those} 
by Rucker and Erickson \cite{Rucker:2024} 
\textcolor{black}{and by Fukushima, Masayuki and Hirahara}; 
\textcolor{black}{both} 
used physics-informed neural networks,
\textcolor{black}{the former} 
to solve the elastodynamics problem (forward and inverse) in a domain that includes an interface governed by a customary RSF law, 
\textcolor{black}{and the latter to perform forward and inverse of aseismic slip}.  
\textcolor{black}{\st{h}}
\textcolor{black}{H}erein, machine learning is not deployed either to improve an empirical approximate law or to solve a larger problem that features one, but to replace it altogether; down the road, we envision neural-network frictional models being embedded in larger-scale simulations\cite{Lapusta:2000,Lapusta:2003,Cattania:2014}. 
It also differs from the use made by Rouet-Leduc \textit{et al} \cite{Rouet-Leduc:2018} and Hulbert  \textit{et al} \cite{Hulbert:2019}, among others \cite{Wang:2022,Karim:2023}, who used machine learning techniques to devise impending rupture predictions based on acoustic emissions recorded in lab settings. 
To the best of our knowledge, the closest work may be the one by Wang and Sun\cite{Wang:2021}, which focused on interfacial traction-opening laws, and the recent one by Cravero and K. Bhattacharya \cite{Cravero:2023}, who studied interface adhesion.  
Herein, we assume a virtual interface that is perfectly modeled by a customary aging or slip law, \cref{eq:rsf_laws} with either \cref{,eq:state_evolution_aging} or \cref{eq:state_evolution_slip}, for which data for training and verification can be generated easily from those equations. 
We use those data to train a GRU network with a custom loss function that learns the dynamics without explicitly defining internal variables \cite{Mohr:2021}, 
since internal variables cannot be accessed experimentally, cf. work by Kumar and collaborators \cite{Sid:2022,Sid:2023}. 

In reality, the complexities of interface physics \cite{Vakis:2018}, depicted schematically in \Cref{fig:scheme}(a) and (b), are such that probably no single internal variable can suffice as long as one aims at capturing any ambitious degree of complexity. 
We presume implicitly that sliding amount and sliding rate are the sole driving forces of the phenomenon, and whatever internal variables there may be can be codified within the hidden states of the net. 
The figure of merit will not be the dynamic friction coefficient itself, but \textit{variations in the friction coefficient} due to sliding. 
This decision is motivated by the practical inability to surmise the time evolution of the friction coefficient except if proper initial conditions are supplied. 
Hence, when using the trained network to make predictions, one should picture an ongoing experiment that has reached a certain steady-state friction level: that information and the subsequent ``slide-hold-slide'' velocity protocol are passed to the network, which outputs a prediction for the friction evolution for the rest of the experiment. 



\begin{figure}
\centering
\captionsetup{labelformat=empty} 
\caption{\Large\textbf{(a)}  \hspace{3cm}   \textbf{(b)}}
\includegraphics[width=0.45\linewidth]{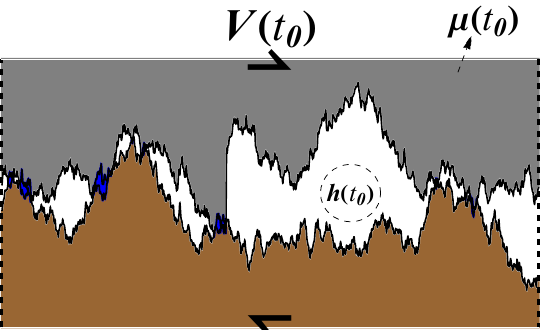}
\hspace{0.5cm}
\includegraphics[width=0.45\linewidth]{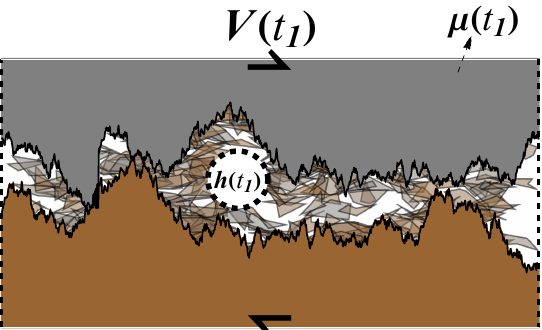}
\includegraphics[width=0.95\linewidth]{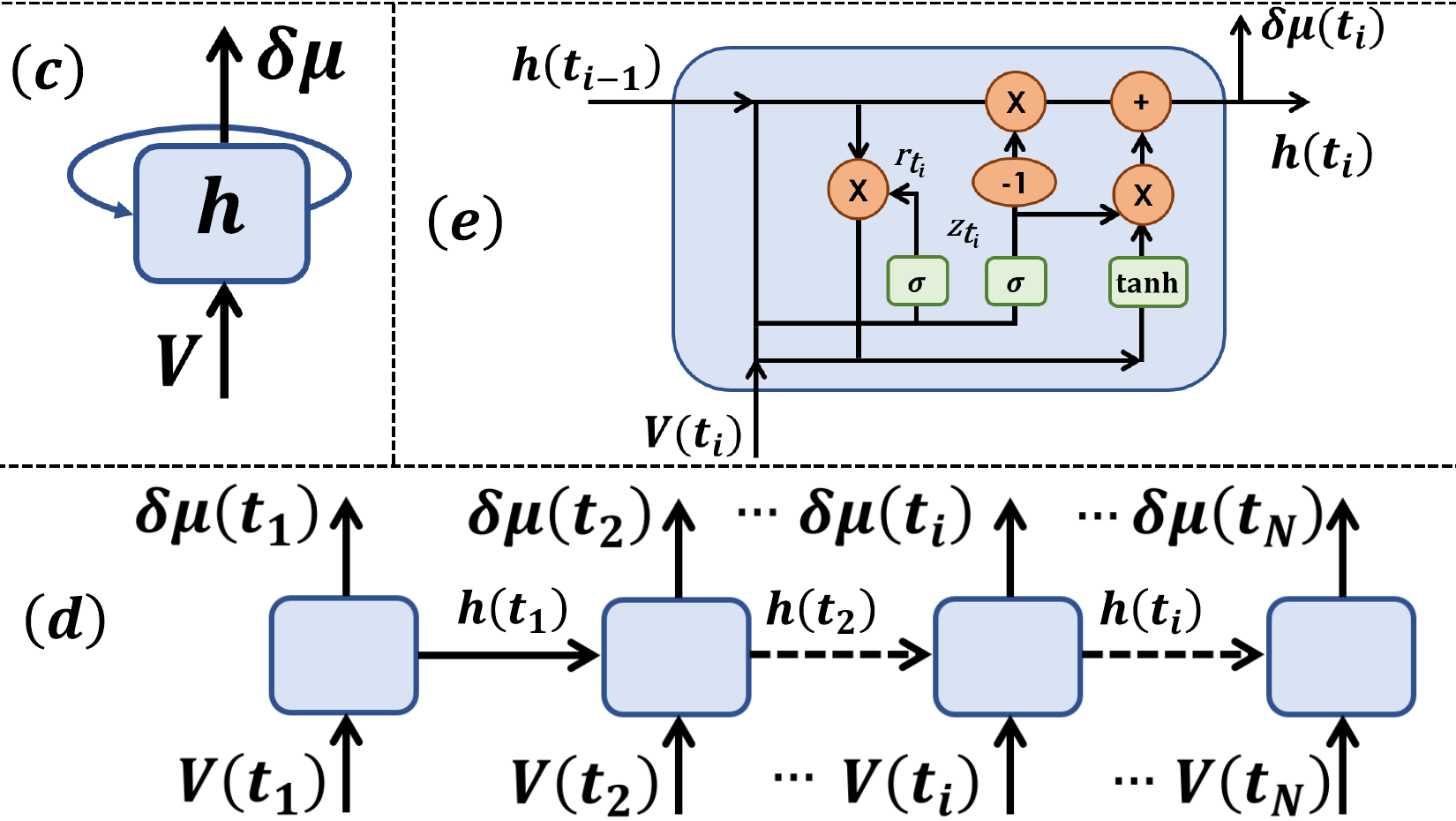}
\setcounter{figure}{0}
\captionsetup{labelformat=default}
\caption{Why using recurrent neural networks to model interface evolution? 
(a) Sketch of the rough interface at time $t_0$. The geometrical and thermomechanical configuration at $t_0$ is assumed to be described by a vector of ``hidden state variables'' evaluated at that time, $\boldsymbol{h}(t_0)$. 
In the usual RSF models, \cref{eq:state_equations}, the hidden state variable would be a scalar, $\theta$, the so-called ``state''. 
A tangential velocity $V(t_0)$ comes imposed remotely.
The ratio between the integrated tangential forces over the vertical ones (all over the interface) yields the instantaneous friction coefficient $\mu (t_0)$, which is a variable that homogenizes the complex ``microscale'' system into a ``macroscale'' description of the resistance to slide one surface over the other, given the velocity at that instant. Thus, $\mu (t_0) = \mu(V(t_0),\boldsymbol{h}(t_0))$. 
(b) As sliding progresses, at time $t_1$ $>$ $t_0$, different micromechanical phenomena (adhesion, creep, plasticity, fracture, wear, heating...) have potentially come into play and modified the configuration of the interface, e.g.,  
the roughness profile has been modified, and debris particles generated by wear are trapped between the surfaces. 
Nevertheless, the current state of the interface can still be described by the hidden variables, $\boldsymbol{h}(t_1)$, 
thus, knowing it along with the current velocity, the overall frictional response of the interface obeys $\mu (t_1) = \mu(V(t_1),\boldsymbol{h}(t_1))$. 
(c) Compact view of the GRU network to represent interface dynamic response. 
The key idea is to replace the simple state variable $\theta$ and either evolution law, \cref{eq:state_equations}, by the hidden state of the RNN $\boldsymbol{h}$, whose evolution is inferred from data. 
The network takes a vector of sliding velocities (``protocol'') $\boldsymbol{V}$ and returns the corresponding variations of friction coefficient $\delta \boldsymbol{\mu }$, while internally handling the interface evolution due to imposed velocities through $\boldsymbol{h}$. 
This depiction is a simplification: between $\boldsymbol{h}$ and $\delta \boldsymbol{\mu}$ there is also a fully-connected network (to ensure proper dimensions).
(d) Deployed view of GRU network. 
The $i$-th cell takes the corresponding velocity and the previous state and returns the new state, $h(t_i)$, which is used to compute the friction evolution at the end. 
(e) Details of gated connections within each cell. The variable $r_{t}$ controls what prior information to forget (``reset'') while $z_{t}$, how to average new and old info, see \cref{eq:gru_mechanism} for all details.
The learning process optimizes the internal parameters that control the variables' values. 
}
\label{fig:scheme}
\end{figure}

\section{Methods}

\subsection{Neural networks for sequences: gated recurrent units}

RNNs are engineered to handle sequential data thanks to their ability to maintain a memory of previous inputs by feeding the output of a neuron back onto itself \cite{Goodfellow:2016}. 
See \Cref{fig:scheme}(c) and (d) for explanatory graphics. 
This recursive nature allows making predictions based on the information accumulated from prior inputs.
However, standard RNNs often struggle with long-term dependencies due to the vanishing gradient problem \cite{Hochreiter:1997}, where backpropagation gradients become too small or too large (over successive passings through a loop) for the network to learn effectively over many time steps \cite{Goodfellow:2016}. 
%
Gated Recurrent Units (GRUs) are an evolution of the traditional RNN architecture, designed to mitigate the vanishing gradient problem and improve the network's ability to learn from long sequences. 
GRUs introduce gated units that control the flow of information,  
what will be useful here to handle evolution of internal variables.
%
The GRU mechanism \cite{Cho:2014} is articulated mathematically through the following variables:

\begin{subequations}
\begin{align}
    \boldsymbol{r}_t &= 
    \sigma(\boldsymbol{W}_{ir} x_t 
    + 
    \boldsymbol{b}_{ir} 
    + 
    \boldsymbol{W}_{hr} \boldsymbol{h}_{t-1} 
    + \boldsymbol{b}_{hr}) 
    \quad &\text{(Reset gate)} \\
    \boldsymbol{z}_t &= 
    \sigma(\boldsymbol{W}_{iz} x_t 
    + \boldsymbol{b}_{iz} 
    + \boldsymbol{W}_{hz} \boldsymbol{h}_{t-1} 
    + \boldsymbol{b}_{hz})
    \quad &\text{(Update gate)} \\
    \boldsymbol{n}_t &= 
    \tanh(\boldsymbol{W}_{in} x_t 
    + \boldsymbol{b}_{in} + 
    \boldsymbol{r}_t \odot (\boldsymbol{W}_{hn} \boldsymbol{h}_{t-1} 
    + \boldsymbol{b}_{hn}))
    \quad &\text{(New candidate gate)} \\
    \boldsymbol{h}_t &= 
    (1 - \boldsymbol{z}_t) \odot \boldsymbol{n}_t 
    + 
    \boldsymbol{z}_t \odot \boldsymbol{h}_{t-1} \, 
    \quad &\text{(Hidden state update)}
\end{align}
\label{eq:gru_mechanism}
\end{subequations}

In this formulation, $x_t \in \mathbb{R}^{d_{in}}$ is the input vector at time step $t$, where $d_{in} = 1$ is the dimensionality of the input (each element in the input sequence is a scalar value of sliding velocity). 
The hidden state $\boldsymbol{h} \in \mathbb{R}^{d_h}$ represents the internal memory of the GRU at time $t$, where $d_h$ is the dimensionality of the hidden state, which can be interpreted as the number of scalar internal variables needed to characterize the interface mechanical and geometrical state at any given time, see \Cref{fig:scheme}(a) and (b). 
The hidden state from the previous time step, $\boldsymbol{h}_{t-1}$, brings past information in the sequence into the computation of the gates and the candidate hidden state.

The reset gate $\boldsymbol{r}_t \in \mathbb{R}^{d_h}$ determines how much of the previous hidden state $\boldsymbol{h}_{t-1}$ should be ``forgotten'' before being used to compute the candidate hidden state $\boldsymbol{n}_t$. 
The update gate $\boldsymbol{z}_t \in \mathbb{R}^{d_h}$ controls the extent to which the previous hidden state $\boldsymbol{h}_{t-1}$ is updated with the new candidate hidden state $\boldsymbol{n}_t$. 
The candidate hidden state $\boldsymbol{n}_t \in \mathbb{R}^{d_h}$ is computed by combining the current input $x_t$ and the reset version of the previous hidden state.

The weight matrices $\boldsymbol{W}_{ir}, \boldsymbol{W}_{iz}, \boldsymbol{W}_{in} \in \mathbb{R}^{d_h \times d_{in}}$ are used to project the input $x_t$ into the space of the reset, update, and new gates, respectively. 
Similarly, the weight matrices $\boldsymbol{W}_{hr}, \boldsymbol{W}_{hz}, \boldsymbol{W}_{hn} \in \mathbb{R}^{d_h \times d_h}$ project the previous hidden state $\boldsymbol{h}_{t-1}$ into these gates. 
The bias vectors $\boldsymbol{b}_{ir}, \boldsymbol{b}_{iz}, \boldsymbol{b}_{in}, \boldsymbol{b}_{hr}, \boldsymbol{b}_{hz}, \boldsymbol{b}_{hn} \in \mathbb{R}^{d_h}$ provide offsets in these linear transformations and extra ``degrees of freedom'' to be adjusted while learning.
The activation functions used are the sigmoid function $\sigma(\cdot)$, which maps the input to a range between 0 and 1, and the hyperbolic tangent function $\tanh(\cdot)$, which maps the input to a range between -1 and 1. The Hadamard product $\odot$ denotes elementwise multiplication.

At the initial time step $t=0$, the hidden state $\boldsymbol{h}_0$ is typically initialized to a vector of zeros. 
During training, the weights and biases are optimized using backpropagation \cite{Goodfellow:2016} to minimize the difference between observations and predictions, given some physical constraints that dynamic friction is known to respect, as codified in the loss function. 
The key idea is to replace the state variable $\theta$ by the hidden state $\boldsymbol{h}$. 
%
%
By doing this, we also ``\textit{outsource}'' the interface evolution modeling to the ``learning'' process (training phase): 
among all the possible interface evolutions that can be parametrized by the network (through giving values to its weights and biases), the one that most closely follows the observations contained in the training data is the one automatically learned through the optimization process, in which parameters are iteratively evolved to better reproduce outputs (under some physical constraints). 
Thus, there is no need for a state evolution equation anymore; once trained, the network can solve the complete evolution given a sliding protocol. 
%
%
Since a black-box approach is adopted \cite{Murdoch:2019} for now, a 
physical interpretation of the hidden state variables ($\boldsymbol{h}$) needs not be ventured at this point.

\subsection{Synthetic data generation}

It is assumed that the state $\theta(t)$ will not be needed by the neural network that is to replace the traditional RSF law; the internal interface evolution will be handled by $\boldsymbol{h}$. 
This means that there will not be vectors of values of $\theta$ being used during training, but it does not imply that \cref{eq:state_evolution_aging} or \cref{eq:state_evolution_slip} are not being used during the datasets' generation. 
Let us reiterate that we assume the hypothetical interface that is perfectly described by \cref{eq:rsf_laws} along with \cref{eq:state_evolution_aging} or \cref{eq:state_evolution_slip}, 
but this fact would be unbeknownst to us if we were to analyze experimental data coming out of that interface. 
Having access directly to the friction coefficient is common to gouge experiments, yet not so much in other settings (e.g., PMMA friction experiments) which display more complex dynamics as stick-slip. 
The stick-slip phenomenon will be addressed in future work.  

So, to generate the dataset, we start by defining a traditional RSF model with specific parameter values, grounded in the fundamental equations presented earlier, with parameter values set as follows:

\begin{itemize}
    \item Friction coefficient, $\mu_0 = 0.5$ (unitless)
    \item Rate-state parameter, $b = 0.015$ (unitless)
    \item Direct effect parameter, $a = 0.005$ (unitless)
    \item Reference velocity, $V_{\text{ref}} = 1.0 \cdot 10^{-5}$ m/s (also taken to be the characteristic value of sliding velocities)
    \item Characteristic slip distance, $D_c = 50.0 \cdot 10^{-6}$ m.
\end{itemize}
These values are inspired by characteristic values reported recently \cite{fryer:2024a,fryer:2024b}. 
1000 random sequences are created to reflect diverse frictional sliding scenarios. 
The pioneering experiences with material plasticity learning \cite{Bessa:2019,Rimoli:2021,Mohr:2020} suggested that using random forcing to generate the datasets is advisable. 
We follow in their footsteps by infusing randomness in our slide-hold-slide protocols.

{\color{black}
We normalize the network using the following characteristic values: $ \delta \mu_{ch} = 0.01$ for the friction coefficient change and $V_{ch} = V_{ref} $ for the velocities. 

Each sequence is designed to simulate a total sliding distance of either 20 $D_c$, other tests were run with 10 $D_c$ and $30 D_c$. 
These sliding distances are chosen based on 
the number of velocity jumps within each sequence
(which are randomly determined drawing from a uniform distribution with values $\{3, 4, 5\}$) 
to leave enough space for the transients to develop. 
The duration of each sequence is obtained dividing the total sliding by the characteristic velocity and
it initiates from a non-zero slip rate, with the initial state of the interface corresponding to the steady state, as defined by the RSF laws: $\theta_{ss} = D_c / V(t=0)$.

Velocity jumps within these sequences occur at random intervals, determined by sampling times from a uniform distribution across the interval $\{0, 1\}$, with 0 marking the start and 1 the end of the sliding protocol. 
Each sequence incorporates a hold period where the velocity drops to zero, simulating intervals where sliding stops until the subsequent jump. 
To avoid issues associated to dividing by zero or trying to compute the logarithm of either zero or negative numbers, a small yet finite velocity is assigned to the hold for computations ($2 \cdot 10^{-9}$ m/s for aging law, $2 \cdot 10^{-3}$ m/s for slip law).  \\
To compute the velocities between jumps, we divide the total sliding distance by the time spans between these jumps, ensuring that the velocities remain constant in these intervals. 
Thus, we define the ``velocity protocol'' for each sequence / virtual experiment. \\
Having the velocity protocol, we proceed to calculate the evolution of the state variable $\theta$ using either \cref{eq:state_evolution_aging} or \cref{eq:state_evolution_slip}, and initial conditions $V( t = 0 ) = V_0$ and $\theta (t = 0) = \theta_{ss}$. 
%
With the velocity protocol and the evolving state variable at hand, we can then compute the friction coefficient's evolution over time, as described by the RSF models (\cref{eq:rsf_laws}).
}

{\color{black}
This is how we obtain }
 our random datasets comprising pairs of ``features" and ``targets". 
Features represent the velocity protocol sampled at 250 equally-spaced instants during the duration of each virtual experiment, while targets correspond to \textit{variations with respect to the initial value} of the friction coefficient sequence at these same instants. 
%
%
The reason for working with variations and not with the actual values of friction coefficient 
is that  there is no way to predict the \textit{initial} values of friction based on these datasets, they have to be provided as initial conditions over which the variations are later superimposed. 
There is the implicit assumption (borne by empirical observations) that the variations depend on sliding velocity, but not on the instantaneous magnitude of the friction coefficient.
{\color{black}
}

$15\%$ of the sequences are put aside for verification and
$15\%$ to test the network once it is trained. 
When it comes to adding noise to the targets, we choose Gaussian noise with standard deviation 0.01 and zero mean.
{\color{black}
No noise is introduced to velocity inputs (the velocity protocol thus mimics the commands that would be passed to a test machine.
}

\subsection{Network details}

\noindent The PyTorch python neural networks package \cite{pyTorch:2019} has been used. 
%
%

\subsubsection{Architecture}
 
The net, which is aimed at replacing \cref{eq:rsf_laws,eq:state_equations} and whose parameters must be learned via training, is made up by a single \texttt{GRUNet}, followed by a folded \texttt{Linear} layer to transforms the output of the GRU through a linear transformation. 
For the hidden state, we first tried $|\boldsymbol{h}|=100$ and then reduce to $|\boldsymbol{h}|=10$. 
Reducing $|\boldsymbol{h}|$ simplifies the net by reducing the number of adjustable parameters (in this case, from 31001 to just 401). 
Simpler models are preferred in turn because they can be more effectively trained in a data-scarce scenario (to reflect this, the larger model is trained with 700 sequences while the smaller one with 105). 
The PyTorch implementation in use features ``weight sharing'', meaning that some parameters are reused among gates and connections to lighten the complexity of the gated unit \cite{pytorchGRU2024}.  



\subsubsection{Loss function}
\label{sec:loss_function}

The aphorism ``the Law is a teacher'' is often attributed to Saint Augustine of Hippo, 
an early Christian philosopher. 
In the case of neural networks, ``the loss is the teacher''. 
Inspired by physics-informed neural networks (PINNs) \cite{PINNs}, we devise a loss function that takes advantage of automatic differentiation \cite{Autodiff:2018} and reflects physical principles to guide the optimization of the parameters of the network.

Let $\delta \boldsymbol{\mu}$ and $\delta \boldsymbol{\hat{\mu}}$ represent the vector of normalized predictions and targets (respectively) for a given vector of features (velocity protocol) $\boldsymbol{V}$, i.e., $\delta \boldsymbol{\mu} = \mathsf{NN}(\boldsymbol{V}/V_{ch})$. 
See that $\delta \boldsymbol{\mu}, \, \delta \boldsymbol{\hat{\mu}}, \, \boldsymbol{V} \in \mathbb{R}^N$ (and $N=250$ in this case). 
The first part of the loss is the mean-absolute error (MAE):
\begin{align}
    \mathsf{loss}_1
    =
    {
    ||\delta \boldsymbol{\mu} - \delta \boldsymbol{\hat{\mu}}||_1
    \over
    N
    } \, ,
    \label{eq:MAE}
\end{align}
where $||\cdot||_1$ represents the $\ell_1$-norm of the vector. 
This term leads the parameter adjustment during training to follow the available data as closely as possible (``supervised learning''). 
The extra loss terms are aimed to choose among the many ways that there are to minimize this average error. 

The second part of the loss represents the difference between the initial values of predictions and targets, 
\begin{align}
    \label{eq:loss2}
    \mathsf{loss}_2
    =
    \left| 
    \delta \mu_1 - \delta \hat{\mu}_1
    \right| 
    = 
    \left| 
    \delta \mu_1 
    \right|
    \, ,
\end{align}
where the sub-index $1$ refers to the first element of every sequence. 
This term is useful to stabilize the first predictions. 
In the same spirit, we also enforce the initial gradient to be zero: 
\begin{align}
    \label{eq:loss3}
    \mathsf{loss}_3
    =
    {\partial (\delta \mu_1) \over \partial V}
    \, ,
\end{align}
since the velocity should not change in the first step. \\
The conditions that experimental evidence impose over the friction coefficient are the direct effect, i.e., logarithmic dependence of velocity magnitude, and healing, i.e., logarithmic increase of the static friction coefficient with time (during holds)

\begin{align} 
\label{eq:log_variation}
    { \partial \mu
    \over
     \partial \log V}
     = 
     \mathrm{c}_1 \, , 
     \quad
     { \partial \mu_s
    \over
     \partial \log t}
     = 
     \mathrm{c}_2 \, , 
\end{align}
where $\mu_s$ is the static friction coefficient, $c_1$ and $c_2$ are interface-dependent constants. 
According to \cref{eq:rsf_laws}, direct effect and frictional healing are controlled by $a$ and $b$, but the fact that these parameters are behind the interface evolution is not known, so the challenge is how to enforce the \textit{existence} of those constants without directly enforcing any particular \textit{value}. 
%
%
We note that the logarithmic dependence implicit in \cref{eq:log_variation} is equivalent to 
\begin{align}
    \Delta \left[
    V {\partial \mu \over \partial V }        
    \right]_{t_1,t_2}
    =
    \left( 
    \left[ 
        V {\partial \mu \over \partial V }
    \right]_{t_1}
    -
    \left[ 
        V {\partial \mu \over \partial V }
    \right]_{t_2}
    \right)^2 
    =
    0 \, ,
\end{align}
for any two successive instants $t_1$ and $t_2$, the derivatives at either time can be efficiently computed using the automatic differentiation capabilities of PyTorch. 
Assuming that the training data has enough resolution to capture every velocity jump, we can safely choose $t_1$ and $t_2$ to be any two consecutive points, and we must have that the logarithmic increment to be zero to ensure that the variations of the friction coefficient with respect to velocity happens in logarithmic fashion. 
Since that should be the case for every two successive data points, the loss term for direct effect must be
\begin{align}
    \label{eq:loss4}
    \mathsf{loss}_4
    =
    \sum_{i = 1}^{N-1}
    \Delta \left[
    V {\partial (\delta \mu) \over \partial V }        
    \right]_{t_i,t_{i+1}}
     \, .
\end{align} 
The terms corresponding to holds should not enter the previous sum, as there the dominant phenomenon is not the direct effect but healing. 
Conveniently, the fact that $V \approx 0$ during holds automatically acts as a ``mask'' \cite{Goodfellow:2016}, effectively removing those terms from the loss. 
The term (\ref{eq:loss4}) guides an ``unsupervised'' learning process, in the sense that there are no actual values to match, but \textit{a priori} physical information that the network must respect. \\
Let us address healing. A loss term to guide the logarithmic growth during holds can be devised in analogy to (\ref{eq:loss4}). Nevertheless, (after some tests, see Supplementary Material) simply letting the supervised learning promoted by (\ref{eq:MAE}) happened to be more effective than adding this extra term. 
See the discussion section for more. 
Future work must be occupied with tailoring the architecture, the training or both to capture healing. 

Finally, when dealing with noisy data we also add a standard weight-regularization term \cite{Goodfellow:2016}, $\mathsf{loss}_5$. 
In summary, the overall loss term that guides the optimization of weights and biases during the training process is:

\begin{align}
\label{eq:loss}
    \mathsf{loss}
    =
    \underbrace{\lambda_1
    \mathsf{loss}_1}_{\scriptsize \shortstack{\text{supervised} \\ \text{learning}}}
    +
    \overbrace{\lambda_2 
    \mathsf{loss}_2
    +
    \lambda_3
    \mathsf{loss}_3}^{\scriptsize \shortstack{\text{initial} \\ \text{conditions}}}
     +
    \underbrace{\lambda_4
    \mathsf{loss}_4}_{\scriptsize \shortstack{\text{direct} \\ \text{effect}}}
    +
    \overbrace{\lambda_{5}
    \mathsf{loss}_{5}}^{\scriptsize \shortstack{\text{weight} \\ \text{regularization}}}
    \, ,
\end{align}
$\lambda_1, \, \lambda_2, \, \lambda_3, \, \lambda_4, \, \lambda_5 \in \mathbb{R}$. 
It has been found that $\lambda_1 = 1$, $\lambda_2 = \lambda_3 = 0.1$ and $\lambda_4 = 0.01 $ yield satisfactory results. 
$\lambda_5 = 0$ whenever noise is absent; 
an extra $\ell_2$-regularization term is added when dealing with noisy data, with $\lambda_{5} = 10^{-4}$.  

\section{Results}

\subsection{Training}

\begin{figure}
    \centering
    \includegraphics[width=0.495\linewidth]{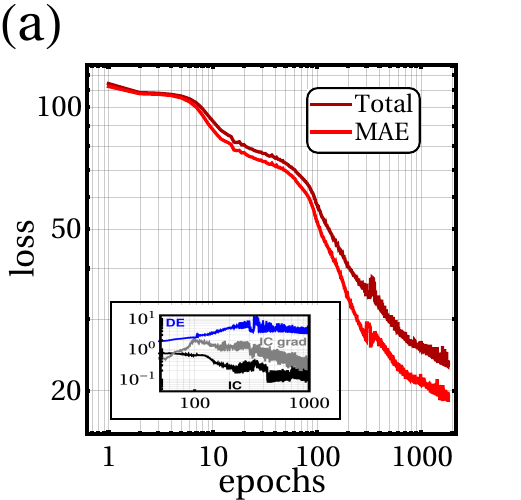}
    \hfill
    \includegraphics[width=0.495\linewidth]{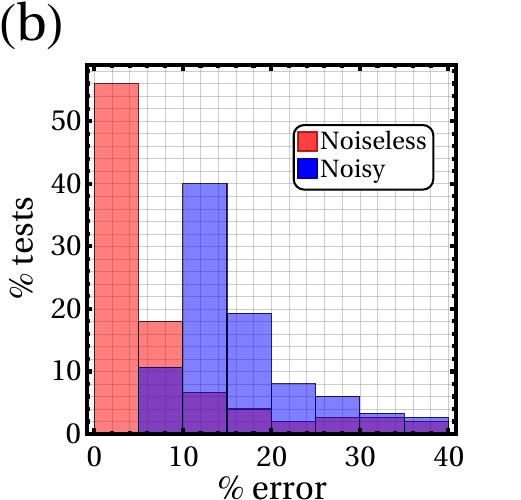}
    \vspace{0.5cm} 
    \vfill    \includegraphics[width=0.495\linewidth]{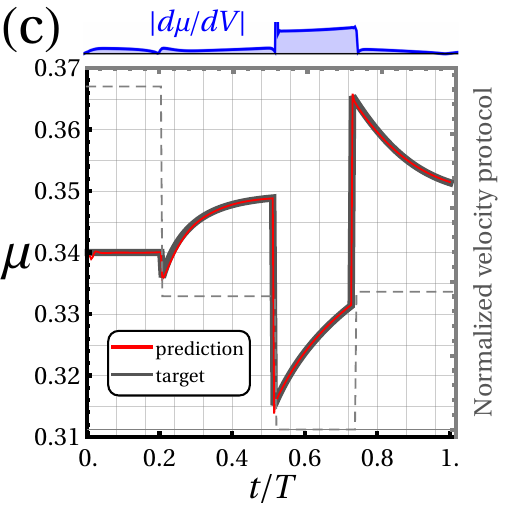}
    \hfill
    \includegraphics[width=0.495\linewidth]{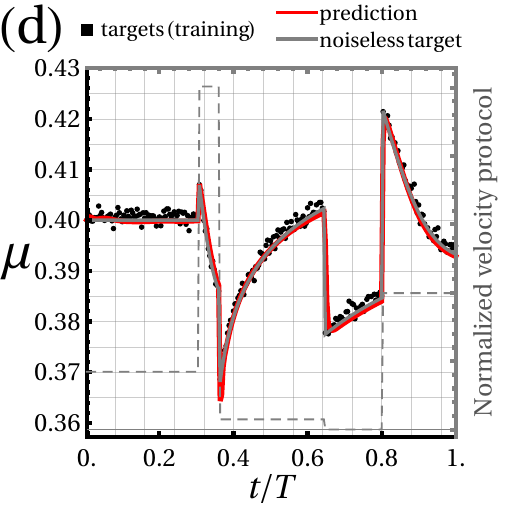}
    \caption{Results sampler. 
    Results correspond to network with hidden state size $|\boldsymbol{h}| = 10$, trained with 105 sequences, batch size equal to 1, $\lambda_1 = 1$, $\lambda_2 = \lambda_3 = 0.1$ and $\lambda_4 = 0.01 $, $\lambda_5 = 0$ for noiseless data and $\lambda_5 = 10^{-4}$ for noisy.  %
    (a) Training loss evolution: most of the loss magnitude corresponds to data tracking in supervised learning. 
    The other terms (inset) become relevant at the end of the process, once parameters had been adjusted to minimize (\ref{eq:MAE}). 
    (b) Error bins (150 test sequences): for noiseless data, the average error is 12\%, while the median error is just 4\%; conversely, for noisy data the average is 17\% and the median 11\%. 
    The error for noisy is computed using the underlying noiseless data. 
    (c) Test example \#1 (noiseless): light right axis for loading velocity, dark left axis for friction coefficient evolution. The dashed background line represents the velocity protocol. Emphasis on gradient magnitude (computed with \texttt{autodiff}): the top inset is aligned with the horizontal normalized time axis ($T$ represents the duration of the experiment), it reveals that most change (intense gradients) happen during the hold and at velocity jumps. 
    (d) Test example \#2 (noisy): similar axes and background as in (c). Emphasis on how the network is able to predict the underlying evolution despite the presence of noise in the training data. 
    }
    \label{fig:results}
\end{figure}

We employ the ADAM optimization method. 
A constant learning rate of 0.001 is chosen. 
We choose to initially use a batch size of 5 when using 700 training sequences for training, then batch sizes of 1 sequence was used to generate the results in \Cref{fig:results}. 
Finally, during training we use a maximum of 2000 epochs combined with early stopping (the training is terminated if the loss does not improve after 100 epochs), 
combined with gradient clipping to avoid erratic exploration of the loss landscape, 
and saving the weights yielding the minimum loss to be the ones used later in inference mode. 

{\color{black}
%
When satisfactory hyperparameters were determined, additional training was carried out on the EPFL's Kuma cluster, on NVIDIA H100 94GB GPUs. 
The purpose of these trainings was to modify the batch size and the number of sequences available for training to find a good balance between the prediction error and the training time. All results are available in \Cref{tab:training_results}. 
When using either 1000 or 150 sequences, we assume that the hidden state vector has 100 components, i.e., $|\boldsymbol{h}|=100$.
As expected, increasing the batch size reduces the neural network's ability to generalize, but also its training time. 
The training time per epoch is inversely proportional to the size of the batch, but the error is not, although it systematically increases with the size of the batch.
For a more concrete application, by indicating an acceptable error of 18.9\% instead of 15.3\% for 1000 sequences, for example, the training time can be reduced by almost 7x (significant in an energy-saving context).}

\begin{table}[H]
{\color{black}
\centering
\begin{tabular}{c@{\hskip 0.5in}cccccc}
\toprule
& \multicolumn{3}{c}{1000 sequences} & \multicolumn{3}{c}{150 sequences} \\
\cmidrule(lr){2-4} \cmidrule(lr){5-7}
Batch size & 5 & 25 & 100 & 5 & 25 & 100 \\
\midrule
CPU/GPU & GPU & GPU & GPU & GPU & GPU & GPU \\
Training time [s] & 17524 & 7331 & 2423 & 4397 & 1867 & 595 \\
Epochs & 683 & 1452 & 1808 & 1118 & 2000 & 1572 \\
Time/Epoch [s] & 25.66 & 5.05 & 1.34 & 3.93 & 0.93 & 0.38 \\
\midrule
Prediction Error [\%] & 15.3 & 16.4 & 18.9 & 14.9 & 16.4 & 21.9 \\
Validation loss & 0.151 & 0.176 & 0.163 & 0.666 & 0.477 & 0.693 \\
\bottomrule
\end{tabular}
\caption{{\color{black}Training time, prediction error and validation loss on the Kuma cluster for different sequence sizes and batch sizes.
For all trainings, $|\boldsymbol{h}|=100$. The datasets used to generate the data in the table correspond always to the aging law \cref{eq:state_evolution_aging}. 
}
}
\label{tab:training_results}
}
\end{table}

For illustration purposes, the overall training procedure is depicted in \Cref{fig:results}(a) and the average errors are consigned in \Cref{fig:results}(b). 
Results presented correspond to the aging law, but similar trends are observed for the slip law too, see also \Cref{tab:table1}.

\subsection{Test}

Once the training is complete, our $\mathsf{NN}$ function to model the interface evolution is ready to use. 
We test it using the 150 sequences, generated with the aging law \cref{eq:state_evolution_aging}, that no net saw during training. 
For each sequence, the percent error is computed as
\begin{align}
    \mathsf{error}(\%)
    =100
    { || \mathsf{NN}(\boldsymbol{V}) -  \delta \boldsymbol{\hat{\mu}} ||_2
    \over 
    || \delta \boldsymbol{\hat{\mu}}||_2} \, ,
\end{align}
the underlying truth is considered to be the noiseless data (this also helps to verify the lack of overfitting). \\ 
The mean test error is consigned in \Cref{tab:table1}. 
Remarkably, we see similar performance for both slip and aging-controlled interfaces, and 
the performance does not deteriorate significantly due to addition of noise. 
The fact that the median error is much smaller than the average one reveals that the cases corresponding to most error are associated with unrealistic slide-hold protocols, in which either the magnitude of velocity changes is unrealistic (too acute) or the jumps happen too quickly. 
These spurious protocols enter the dataset due to the random generation process thereof.

\begin{table}[H]
\centering
\begin{tabular}{lc@{\hskip 0.5in}cc@{\hskip 0.5in}cc}
\toprule
\multirow{2}{*}{\Large $|\boldsymbol{h}|$} & \multicolumn{2}{c}{Noiseless} & \multicolumn{2}{c}{Noisy} \\
\cmidrule(lr){2-3} \cmidrule(lr){4-5}
 & $Aging$ & $Slip$ & $Aging$ & $Slip$ \\
\midrule
10 & 12 (5) & 15 (11) & 18 (12) & 14 (9) \\
100 & 14 (12) & 5 (3) & 12 (8) & 8 (4) \\
\bottomrule
\end{tabular}
\caption{Percent average and median (between parenthesis) errors for different network sizes \textcolor{black}{(trained in all cases with batch size = 1)} : $|\boldsymbol{h}|=10$ (trained with 105 sequences and tested with 22) and $|\boldsymbol{h}|=100$ (700 and 150). Results for both noiseless and noisy data generated using either the aging or the slip state evolution law.}
\label{tab:table1}
\end{table}

\section{Discussion}

\begin{figure}
    \centering
    \includegraphics[width=0.495\linewidth]{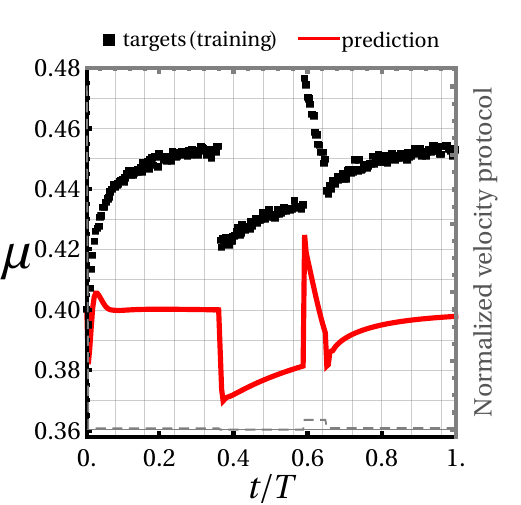}
    \hfill
    \includegraphics[width=0.495\linewidth]{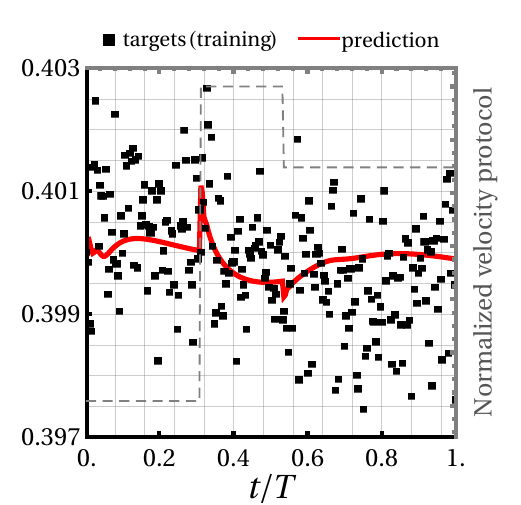}
    \caption{
    \color{black}
    Why does the network fail to predict? 
    Showcasing two representative pathological cases. 
    Results correspond to network with hidden state size $|\boldsymbol{h}| = 100$, trained with 700 sequences, batch size equal to 5, $\lambda_1 = 1$, $\lambda_2 = \lambda_3 = 0.1$ and $\lambda_4 = 0.01 $, $\lambda_5 = 0$ for noiseless data and $\lambda_5 = 10^{-4}$ for noisy.  %
    Left: trong gradient at the beginning of the sequence. 
    The initial velocity is much larger than the one at the next timestep, what leads not only to a large jump magnitude but also to a conflict with the initial condition used to stabilize initial gradients, \cref{eq:loss3}.  
    Right: friction variations subsumed by noise. 
    In this case, the velocity protocol always features relatively low velocities, so much so that the magnitude of the friction changes associated to the velocity jumps are considerably smaller than the amplitude of the noise. 
    }
    \label{fig:fails}
\end{figure}

As indicated by the test results presented in \Cref{tab:table1} and \Cref{fig:results}, the GRU-based networks are able of closely reproducing the friction dynamics encapsulated in the datasets generated with a customary rate-and-state phenomenological law. 
Remarkably, the aforementioned results evince that reasonable amounts of training data ($\sim 100$ experiments) are enough for the network to learn the dynamics, even when noise is present. 
This is a promising result, since the next step in this research effort is to learn from laboratory experiments, which are obviously less numerous than virtual ones. 
We also see in \Cref{fig:results} that the loss term associated to logarithmic velocity dependence, \cref{eq:loss4}, is doing its part in concentrating gradients where the sudden changes in velocity magnitude are detected without compromising the transient evolution after velocity steps. 
The initial conditions, enforced via (\ref{eq:loss2}) and (\ref{eq:loss3}), appear to be properly captured too, regard \Cref{fig:results}(c) and (d). 
 
Nevertheless, there still are obvious limitations. 
For once, there is a need to develop more tailored loss functions to better account for healing since, in practice, experiments do not reveal the evolution of the frictional resistance, the laboratory data shows the relaxation of the experimental setup instead \cite{Scholz:book}. 
As mentioned above, a loss term for unsupervised learning was developed and tested, but a satisfactory training case could not be attained. 
The root cause of this mishap have not been fully elucidated, although it may have to do with the data not featuring ideal logarithmic healing due to the dynamics of the state variable 
(further discussion in this regard is presented in the Supplementary Material). 


{\color{black}

In terms of training itself, a number of factors can be observed and concluded from the results obtained. 
In \Cref{tab:training_results}, GPU trainings have been tested. A GPU cluster was selected because GPUs are generally faster for training large neural networks, due to their ability to perform matrix calculations much more quickly than CPUs. 
However, this capacity is limited by the GPU's memory and reduced precision (16 bits instead of 32 bits). 
In our case, the memory was never exceeded, and the impact of precision on the results was negligible when testing on CPUs and GPUs. 
All other things being equal, using a larger batch size speeds up the training time at the expense of a higher average prediction error, since the parameters of the network are updated less regularly. 
Nevertheless, the difference in error, when comparing to a batch size of 5, is on the order of 5\% for a batch size 20 times larger, which can be assumable in applications that do not necessarily require an extremely precise error but require speedy training. 

The other important point, presented in \Cref{tab:table1}, is the difference between the mean error and the median error. 
Although the mean error may seem high in many cases, the median error is systematically lower. 
This difference is mainly due to pathological cases, in which either (a) the noise subsumes the friction variation, or (b) the randomized velocity test protocols contain extreme values that are not well-represented in the training set. 
\Cref{fig:fails} depicts two such cases. 
%
%
Hence, the median error gives us a slightly more reasonable indication of the performance of our neural network, weighting down the effect of unrepresentative pathological cases.
}

\section{Conclusions}

This document contains an exploratory approach to modeling frictional interface response via deep learning. 
It demonstrates that recurrent neural networks based on Gated Recurrent Units (GRUs) possess the capability to accurately model slide-hold protocols in the context of rate-and-state friction laws, 
provided that the networks are supplied with adequate data and the loss function is tuned accordingly. 
These findings affirm the potential of recurrent neural networks as a promising tool for representing complex frictional behavior and for replacing customary phenomenological dynamic friction laws, while avoiding the need to explicitly define internal variables and ``outsourcing'' the evolution modeling to the training process.  

Looking forward, there are several avenues for enhancing the predictive power and generalizability of recurrent neural networks for dynamic friction. 
Improvement could be sought through the addition of more nuanced physics-informed loss functions predicated on thermodynamic principles \cite{Masi:2021,Chen:2022} and exploration of alternative or \textit{ad hoc} \cite{Bessa:2019,Mohr:2021} RNN architectures. 
Advanced machine learning features such as transformers \cite{attention:2017} and Long Short-Term Memory (LSTM) networks \cite{LSTM:2019} could offer additional improvements. 
These technologies might provide the key to overcoming some of the technical current limitations, e.g., having to work with fixed-length sequences \cite{Li:2021} or lack of ``self-consistency'' \cite{Mohr:2022}. 
But the promise of deep learning does not stop there. 
On one hand, universal models of friction based on minimal state cells \cite{Mohr:2021} could allow modeling different material interfaces with a single numerical tool; 
on the other hand, one can foresee that an approach based on neural \textit{operators} \cite{Li:2021} may help also with discretization dependence, having different data sampling frequencies in the same sequence, and with the problem of generalizing friction from lab conditions to crustal ones. 
%
Incorporating spring-block dynamics and stick-slip \cite{Scholz:book} into the model will be another step for learning from real laboratory data beyond gouge experiments, as e.g. those in PMMA friction.

Despite the optimistic outlook, limitations are still stark. 
Particularly, a pressing challenge for modeling healing is identified, this fundamental aspect of frictional behavior remains to be integrated into the learning scheme. 
Moreover, even though I considered both noiseless and noisy data, the levels of synthetic noise may not be representative of the intrinsic variability that one may expect in laboratory settings, i.e., experiments are much more ``structured'' and seldomly feature the random protocol that have proved so effective when training RNNs. 
Unlike cheap synthetic data, experimental data remains relatively scarce; hence it is worth exploring how new techniques \cite{Yu:2021,Masi:2024}, aimed at improving data economy in the context of constitutive modeling with neural networks, can help in this case. 
It is reassuring for the future steps to know that just dozens of experiments (in lieu of hundreds or thousands of them) may be enough to learn interface dynamics. 
These volumes of information can be generated by a few laboratories, without requiring large-scale collaborations. 
Notwithsatnding that, this paper is also a call to coordination between experimentalists to generate richer datasets that can confidently be used to train models for particular interface conditions. 

The end goal is to evolve from traditional phenomenological friction models towards more accurate representations that are likewise derived from the data already used for parameter fitting. 
This evolution does not entail giving up on a deeper understanding of friction dynamics. 
Rather, it calls for a complementary integration of solid physical principles with the deep-learning paradigm. 
The rationale for adopting these deep-learning models shall be their demonstrable ability to accurately capture complex interface physics, which, in turn, would justify their use in larger-scale numerical simulations involving those phenomena.

\section{Code Availability}

The data necessary to reproduce the results in this paper as well as the script used to generate it and post-process it are publicly available on a Zenodo repository \cite{zenodo} under license Creative Commons Attribution Share Alike 4.0 International. \\
The PyTorch project \cite{pyTorch:2019} is an open-source software maintained by the community and hosted at \texttt{github.com/pytorch/pytorch}. 
Other open-source Python libraries used in this project include Pandas (\texttt{github.com/pandas-dev/pandas}), NumPy (\texttt{github.com/numpy/numpy}) and Matplotlib (\texttt{github.com/matplotlib/matplotlib}). 
Mathematica \cite{Mathematica} is a proprietary software system for symbolic and numerical mathematical computation, developed and licensed by Wolfram Research.









\acknowledgments

The author gratefully acknowledges financial support from the Swiss National Science Foundation, via Ambizione Grant 216341 ``Data-Driven Computational Friction''.
Useful conversations (regarding text improvements, rate-and-state friction and/or laboratory experiments) with Dr. Barnaby Fryer, Dr. Gabriel Meyer, Dr. Simon Guérin-Marthe and Ms. Roxane Ferry are gratefully acknowledged. 
An EPFL bachelor project conducted by Mr. Antoine Binggeli developed Tensorflow code that inspired the use of GRUs to this particular problem. 
%

\noindent \textbf{Declaration of generative AI and AI-assisted technologies in the writing process}. During the preparation of this work the author used GPT4o in order to improve readability. After using this tool, the author reviewed and edited the content as needed and takes full responsibility for the content of the published article.


%
\bibliography{agusample} 
%




%
%
%
%
%

\newpage

\setcounter{page}{1}
\setcounter{figure}{0}
\setcounter{table}{0}

\appendix

\renewcommand{\thefigure}{\Alph{figure}}
\renewcommand{\thetable}{\Alph{table}}

\resetlinenumber[1] 

\begin{center}
    \large \textbf{Data-driven Dynamic Friction Models\\ based on Recurrent Neural Networks} \\
    \vspace{0.5cm}
    \normalsize 
    \textbf{\color{black} Gaëtan Cortes}, 
    \textbf{Joaquin Garcia-Suarez} \\
    \textit{ 
Institute of Civil Engineering,
            \'{E}cole Polytechnique F\'{e}d\'{e}rale de Lausanne (EPFL),\\ CH 1015 Lausanne, 
            Switzerland} \\
    \texttt{\color{black} gaetan.cortes@epfl.ch}, 
    \texttt{joaquin.garciasuarez@epfl.ch} \\
\end{center}


\section*{Supplementary Material}

\subsection{Network Optimization Procedure}

\textcolor{black}{All references to equations follow the numbering in the main body of the paper.}

In the following discussion, all the sequences mentioned have 250 entries, unless explicitly stated otherwise. 
The original GRU network was defined by a hidden state $\boldsymbol{h} \in \mathbb{R}^{100}$ (i.e, there are 100 free parameters to be adjusted during training) and it was trained with 700 training sequences and loss function defined by $\lambda_1 = 1$, $\lambda_2 = \lambda_3 = 0.1$ while taking $\lambda_4$ (direct-effect loss term) equal to 1, 0.1 and 0.01. 
The results in \cref{tab:SMtable1} suggested that $\lambda_4 = 0.01$ rendered the most convenient loss function. 
The values of $\lambda_2$ and $\lambda_3$ were not parametrized as they appeared less influential and their aim (i.e., realizing $\delta \mu_1 = 0$ and $\partial (\delta \mu) / \partial V |_1 = 0$) was fulfilled for this initial value.

\begin{table}[H]
\begin{tabular}{l|ccc}
\hfill $\lambda_4$ (direct-effect penalty) & $1$ & $0.1$ & $0.01$  \\
\hline
Average Test Error (\%) & 49.2 & 7.4 & 5.8  \\
\end{tabular}
\caption{Using the aging law and noiseless data: test error for different $\lambda_4$ ($|\boldsymbol{h}|=100$, 700 training sequences, 150 test sequences, 5 sequences per batch).}
\label{tab:SMtable1}
\end{table}

Once the loss function was defined, it was illustrative to see how the length of the feature and target sequences affected the outcome of the training, all other things left the same. 
We acknowledge (\Cref{tab:SMtable2}) that the test error remains stable when reducing the history length from 250 to 100, but it doubles when doubling from 250 to 500, what suggests that the hidden state would have to be enriched with more free parameters in order to capture long protocols. 
As mentioned in the conclusion of the paper, this illustrated the need to explore neural operators in lieu of neural networks, as they can deal with this issue more naturally. 

\begin{table}[H]
\begin{tabular}{l|ccc}
\hfill sequence length & $100$ & $250$ & $500$ \\
\hline
Average Test Error (\%) & 5.0 & 5.8 & 10.3  \\
\end{tabular}
\caption{ Using the aging law and noiseless data: test error for sequence lengths ($|\boldsymbol{h}|=100$, $\lambda_4 = 0.01$, 700 training sequences, 150 test sequences, 5 sequences per batch).}
\label{tab:SMtable2}
\end{table}

Having the loss function fully specified, we moved to assess the effect of the size of the hidden state, the number of free parameters in the GRU to be adjusted during training, see \cref{tab:SMtable3}. 
It was verified that $\boldsymbol{h} \in \mathbb{R}^{100}$ offered minor improvements over $\boldsymbol{h} \in \mathbb{R}^{50}$, and that $\boldsymbol{h} \in \mathbb{R}^{10}$, despite almost doubling the error, was still below 10\%, the envisioned limit value. 

\begin{table}[H]
\begin{tabular}{l|cccc}
\hfill $|\boldsymbol{h}|$ (hidden-state size) & $1$ & $10$ & $50$ & $100$ \\
\hline
Average Test Error (\%) & 64.8 & 9.6 & 5.9 & 5.8  \\
\end{tabular}
\caption{Using the aging law and noiseless data: test error for different $|\boldsymbol{h}|$  ($\lambda_4 = 0.01$, 700 training sequences, 150 test sequences, 5 sequences per batch).}
\label{tab:SMtable3}
\end{table}

Given that the real data can come by the dozens of laboratory experimental sequences but hardly by the hundreds, it was imperative to assess how sensitive the training is with respect to data volumes. 
An important adjustment to be made in the case of little available training data is reducing the batch size to one, meaning that the network is backward updated after the pass of each and every sequence. 
\Cref{tab:SMtable4} shows these results, and, remarkably, how they maintain the testing error below 10\% using only 105 sequences if also the batch size is reduced to 1. 

\begin{table}[H]
\begin{tabular}{l|ccccc}
\hfill \# training sequences & $700$ (batch 5) & $140$ & $105$ (batch 5) & $105$ & $70$ \\
\hline
Average Test Error (\%) & 9.6 & 7.6 & 18.0 & 9.5 & 14.6 \\
\end{tabular}
\caption{Using the aging law and noiseless data: test error for different number of training sequences ($\lambda_4 = 0.01$, $|\boldsymbol{h}|=10$, 1 sequence per batch unless stated otherwise).}
\label{tab:SMtable4} 
\end{table}

Finally, the test error for the last model ($|\boldsymbol{h}|=10$, 105 training sequences, 22 test sequences, 1 sequence per batch) without the direct effect loss goes up to 13\%, this confirms the beneficial effect of adding the loss term inspired by the direct effect.

\subsection{Loss Term for Frictional Healing during Holds}

Similarly to what happened with the direct effect, \cref{sec:loss_function} in the main text, the goal is to devise a loss term that guides the network to feature logarithmic increments in friction over time increments elapsed during holds. 
Mathematically, this was stated as (cf. \cref{eq:log_variation} in main text)
\begin{align} 
     { d \mu_s
    \over
     d\log t}
     = 
     \mathrm{c}_2 \, , 
     \nonumber
\end{align}
wherein the constant $\mathrm{c}_2$ could be recognized as the RSF parameter ``$b$'', but our goal is precisely to avoid any presupposition as to its value. 
The previous expression can be manipulated via integration as follows:
\begin{align} 
     \mu_s(t)
     -
     \mu_s(N_h \Delta t)
     =
     \mathrm{c}_2
     \log 
     \left( 
     { t
        \over
     N_h \Delta t}
     \right) 
     \to 
     {
     \mu_s(t)
     -
     \mu_s(N_h \Delta t)
        \over
    \log 
     \left( 
     { t
        \over
     N_h \Delta t}
     \right)
     }
     =
     \mathrm{c}_2
      \, , 
     \nonumber
\end{align}
where the time $t$ has origin $t=0$ when the healing phase starts. 
Assuming that actual information comes in terms of vectors, the healing phase is taken to last for a total time $\Delta t N_h$, i.e., $N_h$ consecutive entries of the vector define the hold (the time elapsed between consecutive entries is therefore $\Delta t$). 
Without ever knowing the value of $\mathrm{c}_2$, it follows that for two consecutive entries (corresponding to time $(n+1)\Delta t$ and $n\Delta t$):
\begin{align*}
    &{
     \mu_s((n+1) \Delta t)
     -
     \mu_s(N_h \Delta t)
        \over
    \log 
     \left( 
     { n+1
        \over
     N_h }
     \right)
     }
     =
     \mathrm{c}_2
     =
     {
     \mu_s(n \Delta t)
     -
     \mu_s(N_h \Delta t)
        \over
    \log 
     \left( 
     { n+1
        \over
     N_h }
     \right)
     } \\
     & \Rightarrow 
     {
     \mu_{n+1}
     -
     \mu_{N_h}
        \over
    \log 
     \left( 
     { n+1
        \over
     N_h }
     \right)
     }
     -
     {
     \mu_{n}
     -
     \mu_{N_h}
        \over
    \log 
     \left( 
     { n+1
        \over
     N_h }
     \right)
     }
     =
     0 \, ,
\end{align*}
where the simplified notation $\mu_{n} = \mu_{s}(n \Delta t)$ has been used. 
If the evolution was perfectly logarithmic over time, each of those ``weighted differences'' (weighted by the proportional logarithmic term) should be zero. 
Thus, we should minimize each and every of these terms during training. 
Finally, replacing the actual values of friction by the relative increment with respect to the initial value, $\mu_n = \mu(T=0) + \delta \mu_n $ (using $T$ to refer to the absolute time, unlike $t$ which is relative time passed from the beginning of the hold), and summing all the terms of differences of weighted differences, 
the healing loss term is:
\begin{align}
    \label{eq:healing}
    \mathsf{loss}_6
    =
    \sum_{n=1}^{N_h - 2}
    \left\{
    {\delta \mu_{n+1} - \delta \mu_{N_h}
    \over
    \log 
    \left(
    {n+1 \over N_h}
    \right)}
    -
    {\delta \mu_n - \delta \mu_{N_h}
    \over
    \log 
    \left(
    {n \over N_h}
    \right)}
    \right\}^2 \, .
    \nonumber
\end{align}

A number of tests were run (for different values of $\lambda_6$) but the addition of this term did not bring about improvements. 
It is believed that this was due to limitations in the synthetic data, 
which, by construction, cannot conform to the ideal logarithmic healing, 
%
%
which entails that the increment of static friction coefficient $\Delta \mu_s$ over a timespan $\Delta t$ abides by
\begin{align}
    \Delta \mu_s \propto {\Delta t \over t} \, ,
\end{align}
where $t$ is the total time elapsed from the arrest. Conversely, if we integrate \cref{eq:state_evolution_aging} and then plug it back into \cref{eq:rsf_laws}, a term depending on the initial value of the state variable appears: 
\begin{align}
    \Delta \mu_s \propto {\Delta t \over \theta(t=0) + t} \, ,
\end{align}
which can interfere with the scaling except if $\theta(t=0) \ll t$. 
The latter condition may be met in real experiments, in which the holds tend to last for much longer than the characteristic age of the interface, but it is certainly not the case here as the holds are always ``short'', i.e., of duration similar magnitude to the magnitude of the state variable.  
%
%
In any case, it does not appear as insurmountable and certainly must be fixed: to work with real data, which does not show the frictional healing in real time but the relaxation of the experimental setup and directly the last value of static friction when sliding is enforced again, an unsupervised learning strategy like the one under discussion must be set in place.  

\vfill

\subsection{Extra examples}

The following correspond to the 10 best sequences and 10 worst sequences in test for a model $|\boldsymbol{h}|$ trained with 105 sequences and batches of 1 sequence. 
Focus in particular in the range of driving velocity magnitude, i.e., left vertical sub-axis.

In the following plots, the initial condition is taken to be $\mu = 0.5$ at the beginning of the experiment.

\begin{figure}
    \centering
    \includegraphics[width=1.0\linewidth]{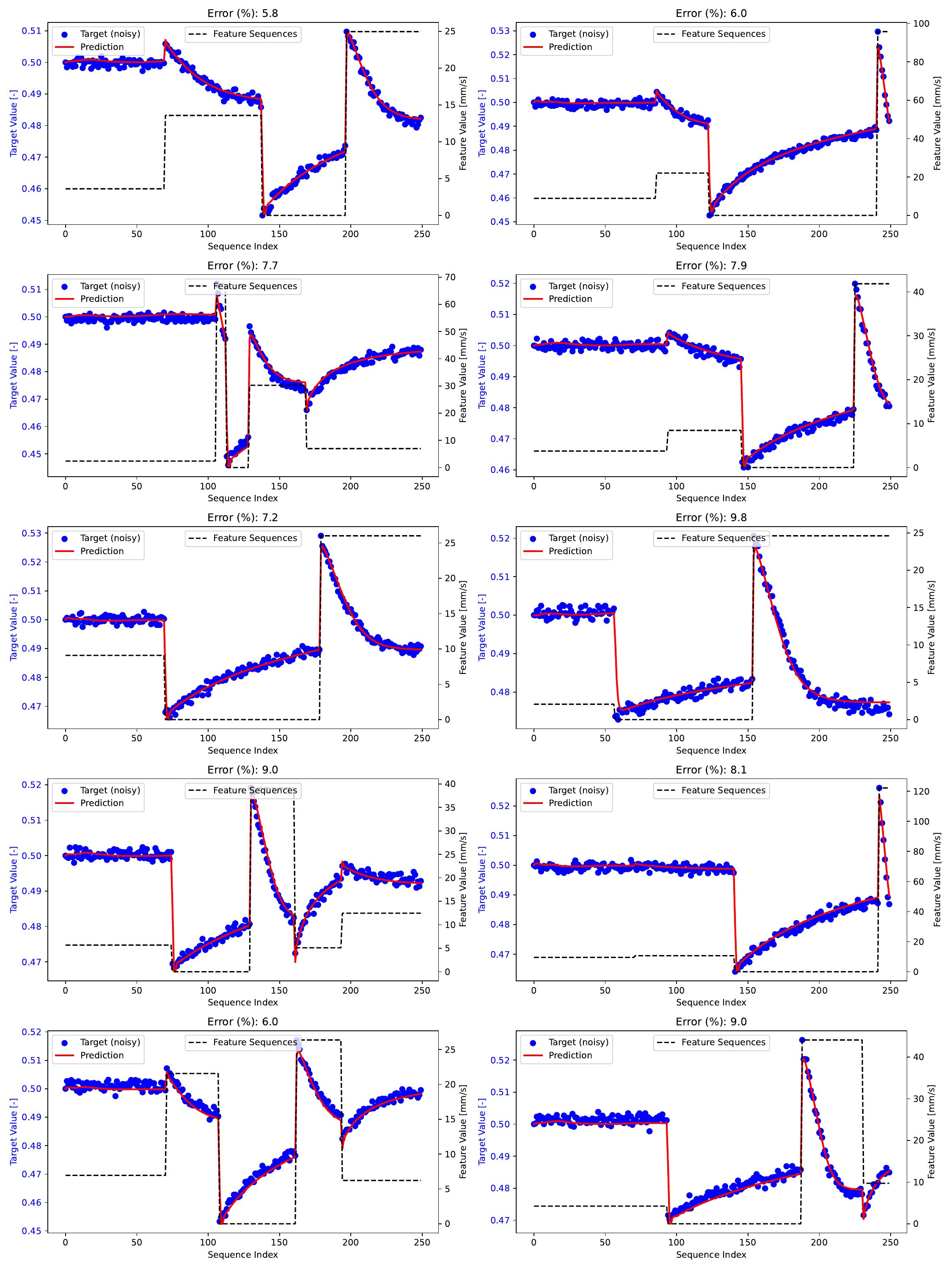}
    \caption{Best performance aging.}
    \label{fig:best_aging}
\end{figure}

\begin{figure}
    \centering
    \includegraphics[width=1.0\linewidth]{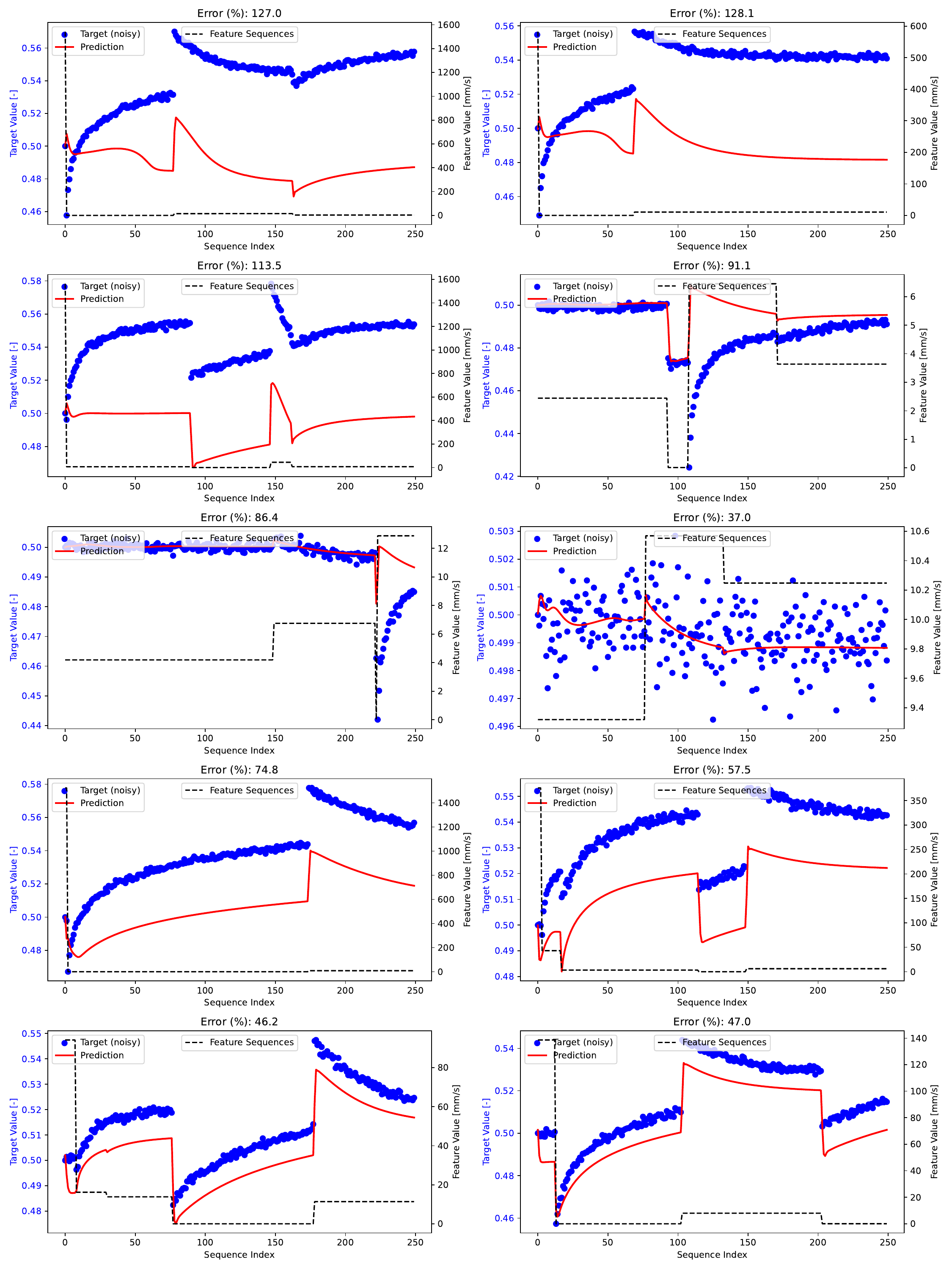}
    \caption{Worst performance aging.}
    \label{fig:worst_aging}
\end{figure}

\begin{figure}
    \centering
    \includegraphics[width=1.0\linewidth]{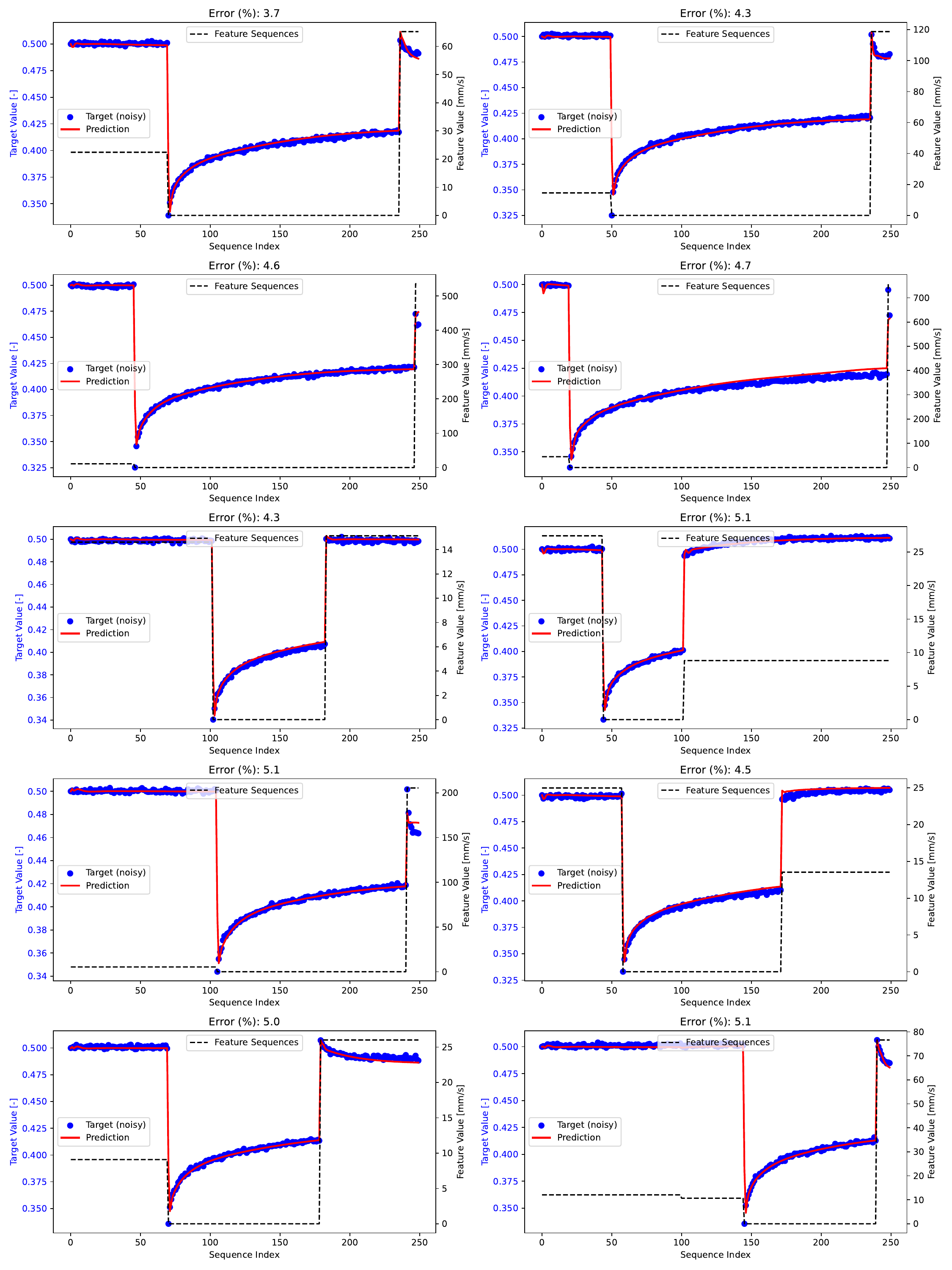}
    \caption{Best performance slip.}
    \label{fig:best_slip}
\end{figure}

\begin{figure}
    \centering
    \includegraphics[width=1.0\linewidth]{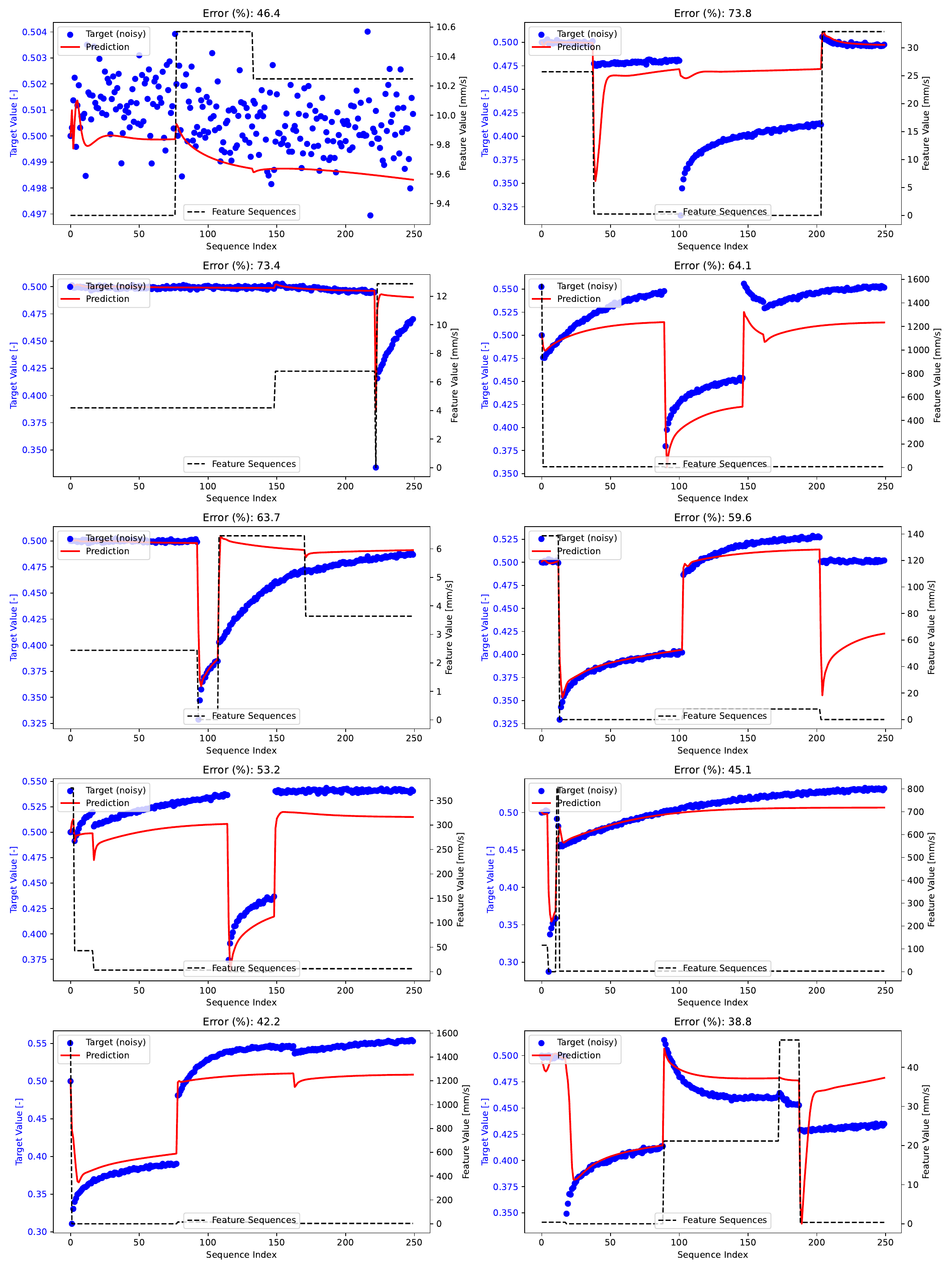}
    \caption{Worst performance slip.}
    \label{fig:worst_slip}
\end{figure}

\end{document}